\documentclass[conference]{IEEEtran}
\IEEEoverridecommandlockouts
\usepackage[T1]{fontenc}
\usepackage{cite}
\usepackage{amsmath,amssymb}
\usepackage{graphicx}
\usepackage{hyperref}
\usepackage{hyphenat}
\usepackage[table,dvipsnames]{xcolor}
\usepackage{multirow}
\usepackage{booktabs}
\usepackage{arydshln}
\usepackage{tcolorbox}
\usepackage{pgfplots}
\usetikzlibrary{plotmarks}
\usepackage{cmll}
\usepackage{etoolbox}
\pgfplotsset{compat=1.18}

\usepackage[skip=1pt]{caption}

\definecolor{Teal}{RGB}{0,128,128}
\definecolor{Peru}{RGB}{205,133,63}

\title{AudioToolAgent: An Agentic Framework for Audio-Language Models}
\author{
\IEEEauthorblockN{Gijs Wijngaard\textsuperscript{1,2,3} \quad Elia Formisano\textsuperscript{1} \quad Michel Dumontier\textsuperscript{1} \quad Jenia Jitsev\textsuperscript{2,3}}
\IEEEauthorblockA{\textsuperscript{1}Maastricht University \quad \textsuperscript{2}LAION \quad \textsuperscript{3}Juelich Supercomputing Center (JSC), Research Center Juelich (FZJ)}
}

\begin{document}

\maketitle

\begin{abstract}
Large Audio-Language Models (LALMs) perform well on audio understanding tasks but lack multistep reasoning and tool-calling found in recent Large Language Models (LLMs). This paper presents AudioToolAgent, a framework that coordinates audio-language models as tools via a central LLM agent that accesses tool adapters for audio question answering and speech-to-text. The agent reasons about which tools to invoke, how to formulate follow-up queries, and how to arbitrate conflicting tool outputs, without accessing the audio. Experiments with MMAU, MMAR, and MMAU-Pro show state-of-the-art accuracy: up to 77.50\% in MMAU, 77.00\% in MMAR, and 61.90\% in MMAU-Pro. Monte Carlo sampling for Shapley values across 490 configurations identifies effective agent-tool combinations. The modular design allows the integration of new tools and removes the need for additional training data and training runs. The code and reproduction materials are available at \url{https://github.com/GLJS/AudioToolAgent}.
\end{abstract}


\section{Introduction}
\label{sec:intro}

Understanding and reasoning about audio is central to human cognition. Recent progress in transferring this capability to machines spans two areas: the advancement of Large Language Models (LLMs) with reasoning and tool-calling capabilities \cite{deepseek-aiDeepSeekV3TechnicalReport2025,teamGLM45AgenticReasoning2025,yangQwen3TechnicalReport2025,openaiteamGPT5SystemCard2025,comaniciGemini25Pushing2025}, and the development of Large Audio-Language Models (LALMs) for tasks such as audio captioning, audio question answering, and speech recognition \cite{xuQwen25OmniTechnicalReport2025,chuQwen2AudioTechnicalReport2024,kimiteamKimiAudioTechnicalReport2025,luDeSTA25AudioGeneralPurposeLarge2025,wuStepAudio2Technical2025,goelAudioFlamingo32025,liuVoxtral2025,radfordRobustSpeechRecognition2022}.

Recent LALMs perform well on audio benchmarks \cite{sakshiMMAUMassiveMultiTask2025,kumarMMAUProChallengingComprehensive2025} but most lack tool use \cite{wuStepAudio2Technical2025}. Conversely, general LLMs excel at reasoning and using external tools but lack direct audio processing. The present work combines these strengths by enabling an LLM agent to use audio models as tools.

The framework introduces \textbf{AudioToolAgent}, which treats audio-language models as tools and uses a central agent to coordinate them. The agent, a text-only LLM, cannot process the audio directly; it receives only the audio file path, a question or prompt, and possible answers. The agent then reasons about which tools to invoke and delegates instructions to LALMs (tools) that process the audio. Performance gains arise from two mechanisms: (i) selective use of complementary tools, combining specialized ASR with general audio understanding models, and (ii) multi-hop reasoning across multiple tools to reduce errors from individual models. Because the framework reuses pretrained models, it requires no new datasets or training.

\begin{figure}[ht]
    \centering
    \includegraphics[width=0.95\linewidth]{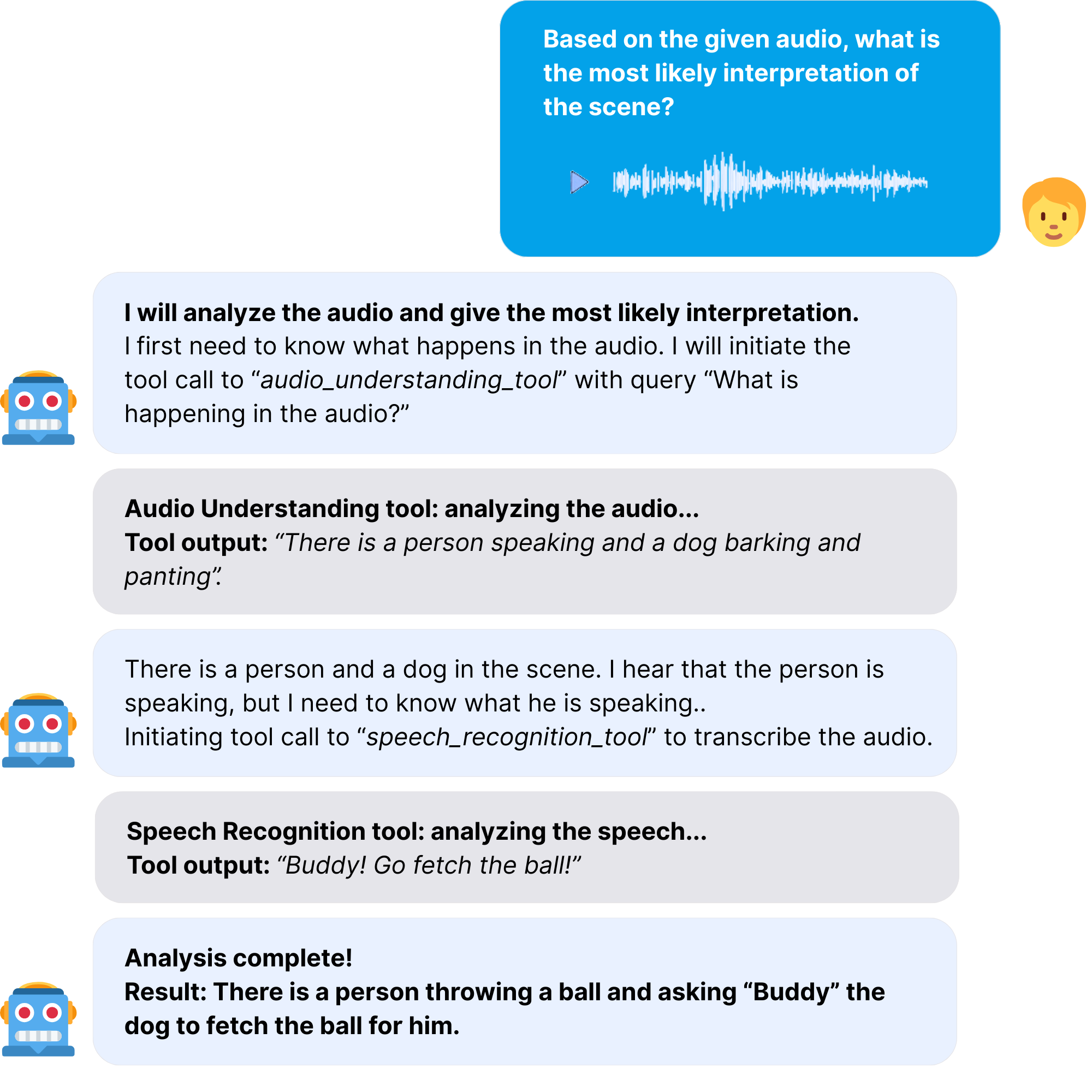}
    \caption{Example of a chatbot using the AudioToolAgent framework. The agent is a large language model that coordinates the tools; the tools are audio-language models.}
    \label{fig:agent:chatbot_vis}
\end{figure}

The agent receives an audio input, a question, and answer choices. The agent uses this information to reason about the task, upon which it calls tools to be able to answer the question or prompt. For speech, the agent prioritizes speech-to-text tools to transcribe the audio. For environmental sounds or music, the agent uses general audio models to gather information. AudioToolAgent asks follow-up questions, iteratively invokes tools, compares outputs, and verifies disagreements by continuing to call tools with different inputs to increase reliability. Figure \ref{fig:agent:chatbot_vis} shows an example of the framework in a chatbot.

The contributions of this work are twofold:
\begin{enumerate}
    \item A reproducible, agent-based coordination framework with modular tool adapters. Using pretrained foundational models without additional data or fine-tuning, this provides a cost-effective approach to high performance on MMAU \cite{sakshiMMAUMassiveMultiTask2025}, MMAR \cite{maMMARChallengingBenchmark2025}, and MMAU-Pro \cite{kumarMMAUProChallengingComprehensive2025}. Both closed-source and open-source versions of AudioToolAgent reach or exceed previously reported results on multiple benchmark subsets.
    \item A controlled empirical study across three benchmarks with a large ablation space (agents and tools) and Shapley-based contribution analysis, establishing a measurement methodology for tool-using audio systems where tools can be swapped as new models appear without retraining.
\end{enumerate}

\section{Related Work}
\label{sec:agent:related}

Recent advances in large language models (LLMs) have resulted in agents who call tools to solve tasks \cite{patilBerkeleyFunctionCalling2025}. This started with the function calling of GPT-3.5 \cite{openaiFunctionCallingOther2023} and includes the Model Context Protocol (MCP) \cite{anthropicIntroducingModelContext2024}, which standardizes interactions with external tools. This work uses the ReAct framework \cite{yaoReActSynergizingReasoning2022}. In ReAct, the agent first reasons about the task and then performs actions to solve it. The agent selects the appropriate tools to answer the question. After receiving tool responses, the agent decides whether to make additional tool calls or answer the question with the information already gathered.

Related work in vision includes Image-of-Thought (IoT) prompting to external tools for a multimodal model to gain information \cite{zhouImageofThoughtPromptingVisual2024}. The ScienceQA model \cite{luLearnExplainMultimodal2022} uses an image captioning model to provide context for a reasoning model. Similarly, SRICE \cite{zhiSeeingReasoningConfidence2025} guides inference by allowing the model to select parts of the image through interactions with external tools.

Recent developments include large multimodal models that integrate audio processing and agentic capabilities within a single architecture. Models such as Gemini 3 \cite{geminiteamNewEraIntelligence2025} and GPT-4o \cite{openaiGPT4oSystemCard2024} handle both audio comprehension and speech recognition while performing tool calls. Training these models costs substantial resources, and they remain closed-source, accessible only through API endpoints.

Another relevant work, StepAudio 2 \cite{wuStepAudio2Technical2025}, capable of audio understanding, received explicit training in tool call and benchmarking on four specific tools: audio search with multimodal RAG, date and time retrieval, weather search, and web search. This model processes audio input, performs tool calls, understands audio content, and generates speech output. Training this integrated model consumed 1.356 trillion tokens over 21 days \cite{wuStepAudio2Technical2025}. In contrast, the AudioToolAgent framework eliminates this training cost by coordinating existing pretrained models.

Similarly, other works with the same name include AudioAgent \cite{anonymousAudioAgentEnhancingTask2024}, which uses audio attributes to optimize prompts using a fine-tuned LLM for audio tools, and the AudioAgent framework \cite{wangAudioAgentLeveragingLLMs2024}, which uses an LLM to orchestrate audio generation and editing.
The current work differs by using a text-only agent that delegates audio understanding to specialized tools without fine-tuning. Instead of using fixed classifiers, AudioToolAgent queries multiple interchangeable tools and cross-checks their outputs for verification.

\section{Methodology and Experimental Setup}
\label{sec:agent:method}

\subsection{Framework Overview}
\label{ssec:agent:framework}

\begin{figure*}[ht]
    \centering
    \includegraphics[width=0.95\textwidth]{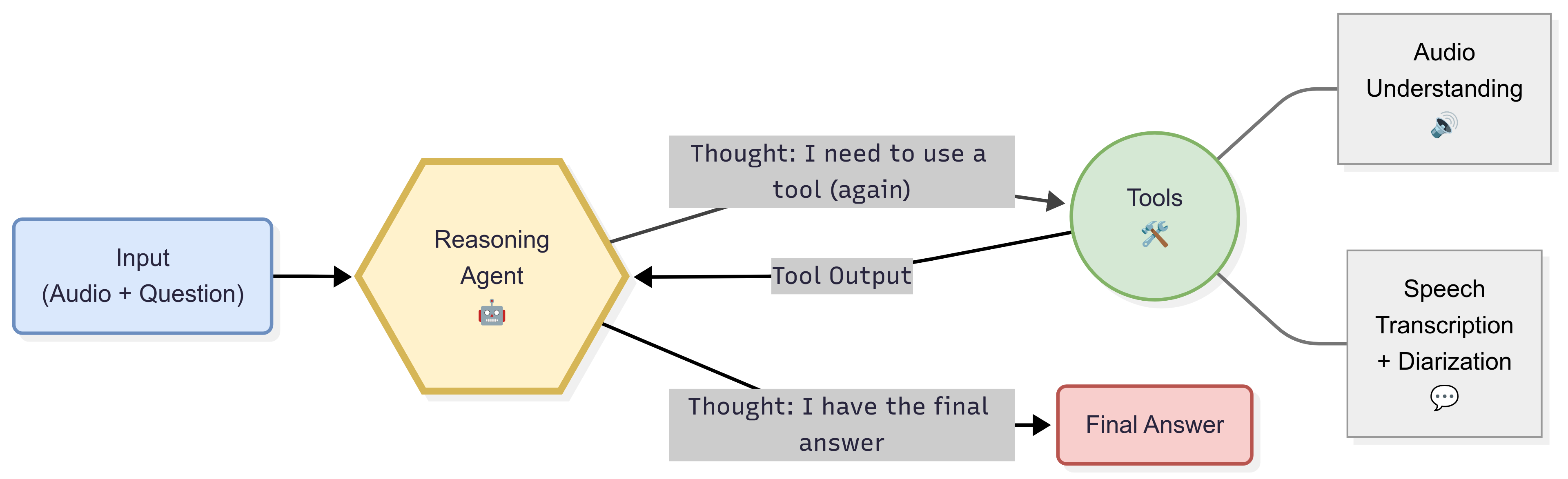}
    \caption{Schematic overview of the AudioToolAgent framework. The framework has two components: a central agent and a set of tool models. The agent is a large language model that coordinates the tools; the tools are audio-language models.}
    \label{fig:agent:audiomcp_diagram}
\end{figure*}

The agent, a reasoning model, receives an audio file path and task description and selects tools to produce the output. The agent accesses audio signals only through the tools. The agent identifies suitable tool calls for the task, then invokes them through structured tags: \texttt{<tool\_call>} to initiate a request and \texttt{</tool\_call>} to conclude it. Within these tags, the agent specifies the target tool, audio file path, and prompt. Each tool's output enters the agent's context, enabling it to reason and invoke additional tools as needed. The set of tools includes multiple speech recognition and audio understanding tools. Figure \ref{fig:agent:audiomcp_diagram} shows a schematic visualization. To prevent runaway loops, each agent can invoke a maximum of 20 tool calls. In AudioToolAgent, the agent makes 5-10 calls, depending on the configuration.

Instructions in the system prompt help the agent to issue follow-up tool calls when outputs conflict or remain ambiguous, and to gather targeted evidence. The prompt also directs the agent to resolve conflicting or ambiguous tool output through targeted follow-up queries and to format answers within \texttt{<answer>} tags for parsing. The system prompt starts as follows:

\begin{tcolorbox}[colback=gray!10,colframe=gray!50,title=System Prompt] \label{box:agent:system-prompt}
You are an expert audio analyst with access to specialized tools. Answer the question given. Put the answer between \texttt{<answer>} and \texttt{</answer>} tags. If the question is multiple choice, there is always just one choice correct. If the tool says it can't listen to audio, try invoking the tool again. Use as many tools as needed to answer the question, even using the same tool multiple times if needed. If initial tool outputs are conflicting or ambiguous, do not guess; instead, you must generate specific, follow-up tool calls to isolate the point of disagreement and gather more detailed evidence. The following tools are available ...
\end{tcolorbox}

In this framework, all tools connect via HTTP API tool adapters for modularity. This includes public endpoints for proprietary models and self-hosted endpoints for open-source models, running on vLLM \cite{kwonEfficientMemoryManagement2023} or Transformers \cite{wolfTransformersStateoftheArtNatural2020}. To demonstrate the versatility of the framework, the implementation offers two configurations, each exposing four tools\footnote{The exact tool registry and configuration files are available at \url{https://github.com/GLJS/AudioToolAgent/}.}:

\begin{itemize}
    \item \textbf{AudioToolAgent}: This configuration uses a proprietary agent and closed-source tools, accessed through public API endpoints to maximize performance. The agent is Gemini 3 Pro \cite{geminiteamNewEraIntelligence2025}, and the tool suite includes Gemini 3 Pro \cite{geminiteamNewEraIntelligence2025}, Qwen2.5 Omni \cite{xuQwen25OmniTechnicalReport2025}, Audio Flamingo 3 \cite{goelAudioFlamingo32025}, and Whisper \cite{radfordRobustSpeechRecognition2022}.
    \item \textbf{AudioToolAgent-Open}: This configuration uses an open-source agent with high-performing open-source audio tools to balance performance with self-hosting capabilities. The agent is Qwen 3 235B \cite{yangQwen3TechnicalReport2025}, and the tool suite includes Qwen3 Omni \cite{yangQwen3TechnicalReport2025}, Qwen2.5 Omni \cite{xuQwen25OmniTechnicalReport2025}, Audio Flamingo 3 \cite{goelAudioFlamingo32025}, and Whisper \cite{radfordRobustSpeechRecognition2022}.
\end{itemize}

The agent uses a multi-hop approach to reduce errors from individual tools. When the agent queries multiple tools and receives conflicting outputs, it triggers targeted re-queries: the agent formulates follow-up questions that constrain the scope to the disputed span or category (e.g., asking specifically about a contested timestamp or sound event). This iterative process continues until the evidence converges across tools or uncertainty remains explicit. By comparing answers across tools, the agent reduces reliance on any single model and mitigates biases inherent in individual tools.

\subsection{Evaluation Setup}
The study evaluated AudioToolAgent and AudioToolAgent-Open on three benchmarks: Massive Multi-Task Audio Understanding (MMAU) \cite{sakshiMMAUMassiveMultiTask2025}, MMAR \cite{maMMARChallengingBenchmark2025} and MMAR-Pro \cite{kumarMMAUProChallengingComprehensive2025}. The MMAU benchmark includes 10,000 audio clips for multi-task audio understanding and reasoning, with 1,000 in the test-mini split and 9,000 in the test split. The experiments used only the test-mini split to reduce costs. The MMAR benchmark tests deep reasoning capabilities with 1,000 audio-question-answer triplets requiring multi-step reasoning across modalities. The MMAU-Pro benchmark measures audio intelligence using 5,305 instances containing question-answer pairs generated by human experts in speech, sound, music, and combinations.

\subsection{Ablation Study Design}
\label{ssec:agent:ablation_design}

To evaluate the effectiveness of different components in the AudioToolAgent framework, an ablation study examines two aspects. First, an agent comparison benchmarks the tool-use performance of state-of-the-art reasoning LLMs while holding the tool set constant. Each agent is evaluated across five independent runs with different random seeds to account for non-deterministic inference behavior. Second, a tool contribution analysis quantifies how much each audio-language model improves performance when added to different tool combinations. The study operates on a 10\% subset (100 examples) of the MMAU test-mini split to balance computational cost with statistical reliability.

For tool analysis, this work uses Shapley values \cite{shapleyValueNPersonGames1953} to quantify whether adding a tool to an existing set of tools is beneficial. Shapley values account for interactions where a tool's value depends on which other tools are present, to be able to calculate marginal contributions in multi-tool systems. To estimate Shapley values, this work uses Monte Carlo sampling on a representative small fraction of all $2^{20} = 1{,}048{,}576$ possible configurations. Without sampling, calculating all Shapley values would require evaluating every possible configuration.

The calculation consists of two stages. In the first stage, the method generates many random permutations of the available tools. For each permutation, the system calculates the accuracy on the benchmark subset of all ordered subsets (coalitions) from the empty set to the full set of tools. In the second stage, the final Shapley value for each tool is calculated across all sampled permutations. This Shapley value is the average of the tool's marginal contribution when added to a pre-existing, non-empty set of tools. Standard errors are computed across permutations to quantify estimation uncertainty.

\section{Results}
\label{sec:agent:results}

\begin{table*}[!ht]
    \centering
    \resizebox{\linewidth}{!}{%
    \begin{tabular}{llc}
    \toprule
    \textbf{Dataset} & \textbf{Models} & \textbf{Results} \\
    \midrule

    \multirow{17}{*}{\shortstack[l]{\textbf{MMAU test-mini} \cite{sakshiMMAUMassiveMultiTask2025} \\
    \textit{Sound $\vert$ Music $\vert$ Speech $\vert$ Average} }} & \multicolumn{2}{l}{\textit{Closed Source}} \\
    & GPT-4o Audio \cite{openaiGPT4oSystemCard2024} & 64.56 $\vert$ 56.29 $\vert$ 66.67 $\vert$ 62.50 \\
    & Gemini 2.5 Flash \cite{comaniciGemini25Pushing2025} & 73.27 $\vert$ 65.57 $\vert$ 76.58 $\vert$ 71.80 \\
    & Gemini 2.5 Pro \cite{comaniciGemini25Pushing2025} & 75.08 $\vert$ 68.26 $\vert$ 71.47 $\vert$ 71.60 \\
    & Gemini 3 Pro \cite{comaniciGemini25Pushing2025} & 71.17 $\vert$ 65.57 $\vert$ 75.08 $\vert$ 70.60 \\
    & Omni-R1$^{\dagger}$ \cite{rouditchenkoOmniR1YouReally2025} & 81.70 $\vert$ 73.40 $\vert$ 76.00 $\vert$ 77.00 \\
    & Step-Audio 2$^{\dagger}$ \cite{wuStepAudio2Technical2025} & \textbf{83.48} $\vert$ \textbf{73.65} $\vert$ 76.88 $\vert$ \textbf{78.00} \\
    \cdashline{2-3}
    \addlinespace[2pt]
    & AudioToolAgent & 81.68 $\vert$ 72.46 $\vert$ \textbf{78.38} $\vert$ 77.50 \\
    \cmidrule(lr){2-3}
    & \multicolumn{2}{l}{\textit{Open Source}} \\
    & Phi-4-multimodal \cite{microsoftPhi4MiniTechnicalReport2025} & 65.47 $\vert$ 64.37 $\vert$ 67.27 $\vert$ 65.70 \\
    & DeSTA2.5-Audio \cite{luDeSTA25AudioGeneralPurposeLarge2025} & 70.27 $\vert$ 56.29 $\vert$ 71.47 $\vert$ 66.00 \\
    & Audio Reasoner \cite{xieAudioReasonerImprovingReasoning2025} & 67.87 $\vert$ 69.16 $\vert$ 66.07 $\vert$ 67.70 \\
    & Kimi-Audio \cite{kimiteamKimiAudioTechnicalReport2025} & 75.68 $\vert$ 66.77 $\vert$ 62.16 $\vert$ 68.20 \\
    & Qwen2.5-Omni \cite{xuQwen25OmniTechnicalReport2025} & 78.10 $\vert$ 65.90 $\vert$ 70.60 $\vert$ 71.50 \\
    & Step-Audio 2 mini \cite{wuStepAudio2Technical2025} & 76.28 $\vert$ \textbf{71.56} $\vert$ 71.47 $\vert$ 73.20 \\
    & Audio Flamingo 3 \cite{goelAudioFlamingo32025} & \textbf{79.58} $\vert$ 66.77 $\vert$ 66.37 $\vert$ 73.30 \\
    \cdashline{2-3}
    \addlinespace[2pt]
    & AudioToolAgent-Open & 78.38 $\vert$ 69.46 $\vert$ \textbf{75.08} $\vert$ \textbf{74.30} \\
    \midrule

    \multirow{12}{*}{\shortstack[l]{\textbf{MMAR} \cite{maMMARChallengingBenchmark2025} \\
    \textit{Sound $\vert$ Music $\vert$ Speech $\vert$ Sound-Music $\vert$} \\ \textit{Sound-Speech $\vert$ Music-Speech $\vert$} \\ \textit{Sound-Music-Speech $\vert$ Average} }} & \multicolumn{2}{l}{\textit{Closed Source}} \\
    & GPT-4o Audio \cite{openaiGPT4oSystemCard2024} & 53.94 $\vert$ 50.97 $\vert$ 70.41 $\vert$ 63.64 $\vert$ 72.48 $\vert$ 62.20 $\vert$ 75.00 $\vert$ 63.50 \\
    & Gemini 2.0 Flash \cite{pichaiIntroducingGemini202024}  & 61.21 $\vert$ 50.97 $\vert$ 72.11 $\vert$ \textbf{81.82} $\vert$ 72.48 $\vert$ 65.85 $\vert$ 70.83 $\vert$ 65.60 \\
    & Gemini 3 Pro \cite{comaniciGemini25Pushing2025} & 72.73 $\vert$ 61.65 $\vert$ 84.35 $\vert$ 72.73 $\vert$ 81.19 $\vert$ 78.05 $\vert$ 83.33 $\vert$ 76.40 \\
    & Omni-R1$^{\dagger}$ \cite{rouditchenkoOmniR1YouReally2025} & 67.30 $\vert$ 51.50 $\vert$ 64.30 $\vert$ 45.50 $\vert$ 70.20 $\vert$ 64.60 $\vert$ 70.80 $\vert$ 63.40 \\
    \cdashline{2-3}
    \addlinespace[2pt]
    & AudioToolAgent & \textbf{73.33} $\vert$ \textbf{63.59} $\vert$ \textbf{82.99} $\vert$ 63.64 $\vert$ \textbf{83.03} $\vert$ \textbf{81.71} $\vert$ \textbf{79.17} $\vert$ \textbf{77.00} \\
    \cmidrule(lr){2-3}
    & \multicolumn{2}{l}{\textit{Open Source}} \\
    & Audio Reasoner \cite{xieAudioReasonerImprovingReasoning2025} & 43.64 $\vert$ 33.50 $\vert$ 32.99 $\vert$ 45.45 $\vert$ 42.66 $\vert$ 31.71 $\vert$ 25.00 $\vert$ 36.80 \\
    & Qwen2.5-Omni \cite{xuQwen25OmniTechnicalReport2025} & 58.79 $\vert$ 40.78 $\vert$ 59.86 $\vert$ \textbf{54.55} $\vert$ 61.93 $\vert$ 67.07 $\vert$ 58.33 $\vert$ 56.70 \\
    \cdashline{2-3}
    \addlinespace[2pt]
    & AudioToolAgent-Open & \textbf{61.21} $\vert$ \textbf{57.28} $\vert$ \textbf{69.05} $\vert$ \textbf{54.55} $\vert$ \textbf{69.27} $\vert$ \textbf{78.05} $\vert$ \textbf{66.67} $\vert$ \textbf{65.90} \\
    \midrule

    \multirow{13}{*}{\shortstack[l]{\textbf{MMAU-Pro} \cite{kumarMMAUProChallengingComprehensive2025} \\
    \textit{Sound $\vert$ Music $\vert$ Speech $\vert$ Sound-Music $\vert$} \\ \textit{Speech-Music $\vert$ Speech-Sound $\vert$} \\
    \textit{Sound-Music-Speech $\vert$ Spatial $\vert$ Voice $\vert$} \\
    \textit{Multi-Audio $\vert$ Open-ended $\vert$} \\
    \textit{Instruction-Following $\vert$ Average} }} & \multicolumn{2}{l}{\textit{Closed Source}} \\
    & GPT4o Audio \cite{openaiGPT4oSystemCard2024} & 44.70 $\vert$ 63.10 $\vert$ 68.20 $\vert$ 40.40 $\vert$ 43.50 $\vert$ 62.50 $\vert$ 57.10 $\vert$ 21.40 $\vert$ 57.50 $\vert$ \textbf{32.60} $\vert$ 43.20 $\vert$ 82.50 $\vert$ 52.50 \\
    & Gemini-2.5 Flash \cite{comaniciGemini25Pushing2025} & 51.90 $\vert$ 64.90 $\vert$ 73.40 $\vert$ 42.80 $\vert$ 58.70 $\vert$ 61.30 $\vert$ 42.80 $\vert$ 36.30 $\vert$ 71.70 $\vert$ 21.20 $\vert$ \textbf{67.50} $\vert$ \textbf{95.10} $\vert$ 59.20 \\
    & Gemini 3 Pro \cite{comaniciGemini25Pushing2025} & 50.76 $\vert$ 69.68 $\vert$ 83.39 $\vert$ 38.00 $\vert$ \textbf{73.91} $\vert$ 73.86 $\vert$ \textbf{71.43} $\vert$ 32.62 $\vert$ 59.66 $\vert$ 24.88 $\vert$ 60.96 $\vert$ 78.16 $\vert$ 60.72 \\
    \cdashline{2-3}
    \addlinespace[2pt]
    & AudioToolAgent & \textbf{54.96} $\vert$ \textbf{72.57} $\vert$ \textbf{87.09} $\vert$ \textbf{52.00} $\vert$ 69.57 $\vert$ \textbf{76.14} $\vert$ 57.14 $\vert$ \textbf{46.15} $\vert$ \textbf{81.03} $\vert$ 30.00 $\vert$ 36.80 $\vert$ 40.23 $\vert$ \textbf{61.90} \\
    \cmidrule(lr){2-3}
    & \multicolumn{2}{l}{\textit{Open Source}} \\
    & Phi4 Multimodal \cite{microsoftPhi4MiniTechnicalReport2025} & 25.70 $\vert$ 47.80 $\vert$ 47.60 $\vert$ 30.00 $\vert$ 39.10 $\vert$ 30.10 $\vert$ 28.60 $\vert$ 39.70 $\vert$ 42.70 $\vert$ 11.40 $\vert$ 42.50 $\vert$ 65.40 $\vert$ 38.70 \\
    & Audio-Reasoner \cite{xieAudioReasonerImprovingReasoning2025} & 34.20 $\vert$ 50.10 $\vert$ 44.00 $\vert$ 26.00 $\vert$ 36.90 $\vert$ 43.20 $\vert$ 28.60 $\vert$ 20.30 $\vert$ 43.40 $\vert$ 22.60 $\vert$ 38.60 $\vert$ 43.40 $\vert$ 39.50 \\
    & DeSTA2.5-Audio \cite{luDeSTA25AudioGeneralPurposeLarge2025} & 35.70 $\vert$ 48.20 $\vert$ 49.90 $\vert$ 22.00 $\vert$ 36.90 $\vert$ 35.20 $\vert$ 28.60 $\vert$ 28.00 $\vert$ 51.00 $\vert$ 19.80 $\vert$ 36.40 $\vert$ 46.50 $\vert$ 40.60 \\
    & Kimi-Audio \cite{kimiteamKimiAudioTechnicalReport2025} & 46.00 $\vert$ 57.60 $\vert$ 52.20 $\vert$ \textbf{46.00} $\vert$ 54.30 $\vert$ 48.90 $\vert$ 42.80 $\vert$ \textbf{43.70} $\vert$ 50.60 $\vert$ 17.20 $\vert$ 34.50 $\vert$ 42.30 $\vert$ 46.60 \\
    & Audio Flamingo 3 \cite{goelAudioFlamingo32025} & \textbf{55.90} $\vert$ 61.70 $\vert$ 58.80 $\vert$ 40.00 $\vert$ 41.30 $\vert$ 47.70 $\vert$ \textbf{57.10} $\vert$ 26.80 $\vert$ 58.60 $\vert$ 26.00 $\vert$ 44.20 $\vert$ 33.30 $\vert$ 51.70 \\
    & Qwen2.5-Omni \cite{xuQwen25OmniTechnicalReport2025} & 47.60 $\vert$ 61.50 $\vert$ 57.40 $\vert$ 40.00 $\vert$ 53.20 $\vert$ \textbf{60.20} $\vert$ 28.50 $\vert$ 41.20 $\vert$ 60.00 $\vert$ 24.30 $\vert$ 52.30 $\vert$ 61.30 $\vert$ 52.20 \\
    \cdashline{2-3}
    \addlinespace[2pt]
    & AudioToolAgent-Open & 46.50 $\vert$ \textbf{65.70} $\vert$ \textbf{72.70} $\vert$ 36.00 $\vert$ \textbf{56.50} $\vert$ 55.70 $\vert$ \textbf{57.10} $\vert$ 42.80 $\vert$ \textbf{66.20} $\vert$ \textbf{27.70} $\vert$ \textbf{62.20} $\vert$ \textbf{58.60} $\vert$ \textbf{57.50} \\
    \bottomrule
    \end{tabular}
    }
    \vspace{2pt}
    \caption{\small Comparison of AudioToolAgent with baseline models on MMAU, MMAR, and MMAU-Pro benchmarks. AudioToolAgent achieves state-of-the-art or near state-of-the-art performance across benchmarks. $^{\dagger}$Self-proposed, no code or API available to verify.}
    \label{tab:main_results}
\end{table*}

Table \ref{tab:main_results} summarizes results across three benchmarks (MMAU, MMAR and MMAU-Pro). AudioToolAgent-Open achieves state-of-the-art open-source performance: 74.30\% on MMAU, 65.90\% on MMAR and 57.50\% on MMAU-Pro, surpassing Audio Flamingo 3, Qwen2.5-Omni, and Kimi-Audio. AudioToolAgent reaches 77.50\% on MMAU, 77.00\% on MMAR and 61.90\% on MMAU-Pro, outperforming GPT-4o Audio and Gemini 2.5 Pro.

Speech tasks show the strongest gains. AudioToolAgent achieves 78.38\% on MMAU Speech and 82.99\% on MMAR Speech. On MMAR speech categories, AudioToolAgent reaches 83.03\% on Sound-Speech and 79.17\% on Sound-Music-Speech, surpassing all baselines. This advantage comes from a specialized ASR tool (Whisper \cite{radfordRobustSpeechRecognition2022}), which provides value even when general omni models already support speech transcription.

ASR models trained with alignment constraints such as CTC, RNN-T, or transducer architectures return tokens only when the acoustics provides sufficient evidence, making them less prone to hallucination than general audio-language models. The ablation results in Section \ref{sec:agent:ablation} confirm that adding dedicated ASR tools improves accuracy on speech-heavy subsets even when omni tools are present.

Beyond speech, both models perform well on cross-domain reasoning and music. On MMAR, AudioToolAgent leads all models on Music-Speech (81.71\%) and Sound-Speech (83.03\%); AudioToolAgent-Open leads all open-source models on every MMAR category. Music understanding is a consistent strength: AudioToolAgent reaches 72.57\% on MMAU-Pro and 72.46\% on MMAU, while AudioToolAgent-Open scores 65.70\% and 69.46\% respectively, the highest open-source results on both benchmarks.

\subsection{Ablation Study}
\label{sec:agent:ablation}

To identify the highest-accuracy configuration for AudioToolAgent, an ablation study examines 10\% of the MMAU test-mini split (100 examples), analyzing agents and tools separately. The methodology for this analysis is also described in Section \ref{ssec:agent:ablation_design}.

\subsubsection{Agents}
The experiments used a fixed set of five open-source tools to evaluate the LLMs capable of tool calling: Qwen2.5 Omni \cite{xuQwen25OmniTechnicalReport2025}, AudioFlamingo 3 \cite{goelAudioFlamingo32025}, DeSTA 2.5 \cite{luDeSTA25AudioGeneralPurposeLarge2025}, Whisper \cite{radfordRobustSpeechRecognition2022}, and Voxtral \cite{liuVoxtral2025}. Each evaluation ran five independent tests per agent with different random seeds and reported the mean. This approach accounts for accuracy variations from non-deterministic inference even with fixed seeds, partly due to vendor-recommended decoding defaults such as nonzero temperature. Figure \ref{fig:model_accuracies} shows the results. Inspired by Omni-R1 \cite{rouditchenkoOmniR1YouReally2025}, which showed that text-only models perform well on audio reasoning tasks, the black vertical tick shows each LLM's no-tools baseline. The dots represent individual evaluations that show discrepancy between results; to combat this, the horizontal colored bar shows the average across 5 runs.

\begin{figure}[ht!]
    \scalebox{0.85}{
    \centering
    \begin{tikzpicture}
        \begin{axis}[
            xbar,
            width=0.85\columnwidth,
            height=8.3cm,
            xmin=0.35, xmax=0.85,
            xlabel={Accuracy},
            ytick={1,2,3,4,5,6,7,8,9,10,11,12},
            yticklabels={
                Mistral Large 3 \cite{mistralteamIntroducingMistral32025},
                Deepseek V3.1 \cite{deepseek-aiDeepSeekV3TechnicalReport2025},
                GPT-5 \cite{openaiteamGPT5SystemCard2025},
                GPT-5 Mini \cite{openaiteamGPT5SystemCard2025},
                Hermes 4 405B \cite{tekniumHermes4Technical2025},
                Kimi K2 \cite{teamKimiK2Open2025},
                Gemini 2.5 Flash \cite{comaniciGemini25Pushing2025},
                Claude Sonnet 4.0 \cite{anthropicteamIntroducingClaude42025},
                Qwen 3 Next 80B A3B \cite{yangQwen3TechnicalReport2025},
                Qwen 3 235B A22B \cite{yangQwen3TechnicalReport2025},
                Gemini 3 Pro \cite{geminiteamNewEraIntelligence2025},
                Claude Opus 4.5 \cite{anthropicteamIntroducingClaude42025},
            },
            y dir=reverse,
            bar width=10pt,
            enlarge y limits=0.06,
            axis x line*=bottom,
            axis y line*=left,
            grid=major,
            grid style={dashed, gray!30},
            tick label style={font=\small},
            label style={font=\small},
            legend style={font=\small, at={(1.0,1.1)}, anchor=north east},
            legend cell align={left},
            legend columns=1,
            every axis plot/.append style={forget plot},
            bar shift auto=false,
            bar shift=0pt,
        ]
            \addplot+[fill=Salmon, draw=none] coordinates {(0.616,1)}; 
            \addplot+[fill=ForestGreen, draw=none] coordinates {(0.656,2)}; 
            \addplot+[fill=RoyalPurple, draw=none] coordinates {(0.658,3)}; 
            \addplot+[fill=Magenta, draw=none] coordinates {(0.670,4)}; 
            \addplot+[fill=Brown, draw=none] coordinates {(0.676,5)}; 
            \addplot+[fill=OliveGreen, draw=none] coordinates {(0.688,6)}; 
            \addplot+[fill=BrickRed, draw=none] coordinates {(0.690,7)}; 
            \addplot+[fill=MidnightBlue, draw=none] coordinates {(0.694,8)}; 
            \addplot+[fill=Orange, draw=none] coordinates {(0.700,9)}; 
            \addplot+[fill=NavyBlue, draw=none] coordinates {(0.710,10)}; 
            \addplot+[fill=Red, draw=none] coordinates {(0.734,11)}; 
            \addplot+[fill=Violet, draw=none] coordinates {(0.752,12)}; 

            \addplot+[only marks, mark=*, mark size=1.5pt, black] coordinates {
                (0.60,1) (0.63,1) (0.62,1-0.06) (0.61,1) (0.62,1+0.06)
            };
            \addplot+[only marks, mark=*, mark size=1.5pt, black] coordinates {
                (0.67,2-0.06) (0.61,2) (0.69,2) (0.64,2) (0.67,2+0.06)
            };
            \addplot+[only marks, mark=*, mark size=1.5pt, black] coordinates {
                (0.68,3) (0.63,3) (0.66,3) (0.65,3) (0.67,3)
            };
            \addplot+[only marks, mark=*, mark size=1.5pt, black] coordinates {
                (0.68,4-0.06) (0.68,4) (0.65,4) (0.66,4) (0.68,4+0.06)
            };
            \addplot+[only marks, mark=*, mark size=1.5pt, black] coordinates {
                (0.68,5) (0.67,5-0.06) (0.67,5) (0.69,5) (0.67,5+0.06)
            };
            \addplot+[only marks, mark=*, mark size=1.5pt, black] coordinates {
                (0.71,6) (0.66,6) (0.69,6) (0.68,6) (0.70,6)
            };
            \addplot+[only marks, mark=*, mark size=1.5pt, black] coordinates {
                (0.69,7) (0.66,7) (0.72,7) (0.68,7) (0.70,7)
            };
            \addplot+[only marks, mark=*, mark size=1.5pt, black] coordinates {
                (0.67,8) (0.68,8) (0.73,8) (0.70,8) (0.69,8)
            };
            \addplot+[only marks, mark=*, mark size=1.5pt, black] coordinates {
                (0.73,9) (0.67,9) (0.69,9) (0.71,9) (0.70,9)
            };
            \addplot+[only marks, mark=*, mark size=1.5pt, black] coordinates {
                (0.71,10-0.06) (0.71,10) (0.70,10) (0.72,10) (0.71,10+0.06)
            };
            \addplot+[only marks, mark=*, mark size=1.5pt, black] coordinates {
                (0.70,11) (0.74,11-0.06) (0.77,11) (0.74,11+0.06) (0.72,11)
            };
            \addplot+[only marks, mark=*, mark size=1.5pt, black] coordinates {
                (0.78,12-0.06) (0.70,12) (0.78,12+0.06) (0.74,12) (0.76,12)
            };

            \draw[black, line width=0.9pt] (axis cs:0.46,1-0.18) -- (axis cs:0.46,1+0.18);
            \draw[black, line width=0.9pt] (axis cs:0.50,2-0.18) -- (axis cs:0.50,2+0.18);
            \draw[black, line width=0.9pt] (axis cs:0.48,3-0.18) -- (axis cs:0.48,3+0.18);
            \draw[black, line width=0.9pt] (axis cs:0.39,4-0.18) -- (axis cs:0.39,4+0.18);
            \draw[black, line width=0.9pt] (axis cs:0.48,5-0.18) -- (axis cs:0.48,5+0.18);
            \draw[black, line width=0.9pt] (axis cs:0.48,6-0.18) -- (axis cs:0.48,6+0.18);
            \draw[black, line width=0.9pt] (axis cs:0.38,7-0.18) -- (axis cs:0.38,7+0.18);
            \draw[black, line width=0.9pt] (axis cs:0.51,8-0.18) -- (axis cs:0.51,8+0.18);
            \draw[black, line width=0.9pt] (axis cs:0.53,9-0.18) -- (axis cs:0.53,9+0.18);
            \draw[black, line width=0.9pt] (axis cs:0.54,10-0.18) -- (axis cs:0.54,10+0.18);
            \draw[black, line width=0.9pt] (axis cs:0.42,11-0.18) -- (axis cs:0.42,11+0.18);
            \draw[black, line width=0.9pt] (axis cs:0.53,12-0.18) -- (axis cs:0.53,12+0.18);

            \pgfplotsset{legend image code/.code={\draw[#1] plot coordinates {(0cm,0cm)};}}
            \addlegendimage{only marks, mark=|, black, mark options={solid, line width=1.0pt}, mark size=3.0pt}
            \addlegendentry{No-tools LLM baseline}
            \addlegendimage{only marks, mark=*, mark size=1.6pt, black}
            \addlegendentry{Single run accuracy}
        \end{axis}
    \end{tikzpicture}
    }
    \caption{Accuracy comparison across 12 LLMs on a subset of the MMAU test-mini split. Bar is an average of 5 runs.}
    \label{fig:model_accuracies}
\end{figure}

Claude Opus 4.5 achieves the highest mean accuracy of 0.752, followed by Gemini 3 Pro (0.734) and Qwen 3 235B (0.710). Closed-source models (Opus 4.5, Gemini 3 Pro) demonstrate the strongest orchestration capabilities, while open-source models like Qwen 3 235B and Qwen 3 Next 80B (0.700) achieve competitive results. All models benefit from tool access, with gains ranging from 15.6 percentage points (Mistral Large 3) to 31.0 percentage points (Gemini 2.5 Flash). Consistency varies: Qwen 3 235B exhibits the narrowest accuracy range (0.02), while DeepSeek V3.1 shows higher variability (range of 0.08). The gap between the weakest agent (Mistral Large 3: 0.616) and the strongest (Opus 4.5: 0.752) spans 13.6 percentage points, demonstrating that orchestration quality is a key differentiator.

Based on these results, AudioToolAgent uses Gemini 3 Pro as the agent, since Claude Opus 4.5 was too expensive. Gemini 3 Pro also serves as an audio tool. AudioToolAgent-Open uses Qwen 3 235B as the highest-performing open-source agent.

\subsubsection{Tools}
This work applies the Shapley values methodology (see Section \ref{ssec:agent:ablation_design}) to quantify each tool's contribution to system performance. Figure \ref{fig:agent:shapley_values} displays the estimated Shapley values with standard error bars computed on 490 sampled configurations across 20 tools. The analysis groups the tools into 4 categories, shown on the y-axis of Figure \ref{fig:agent:shapley_values}.

\begin{figure}[ht]
    \scalebox{0.85}{
    \centering
    \begin{tikzpicture}
        \begin{axis}[
            xbar,
            width=0.85\columnwidth,
            height=10.5cm,
            xlabel={Shapley value},
            xmin=-0.05, xmax=0.15,
            scaled ticks=false,
            xtick scale label code/.code={},
            xtick={-0.05, 0.00, 0.05, 0.10, 0.15},
            xticklabels={-0.05, 0.00, 0.05, 0.10, 0.15},
            ytick={1,2,3,4,5,6,7,8,9,10,11,12,13,14,15,16,17,18,19,20},
            yticklabels={
                Tavily Web Search \cite{TavilyWebAccess},
                DuckDuckGo Web Search \cite{DuckDuckGoProtectionPrivacy},
                Parakeet TDT v3 \cite{cohenNowWereTalking2025},
                Granite Speech \cite{saonGraniteSpeechOpenSourceSpeechAware2025},
                Speaker Diarization \cite{bredinPyannoteaudio21Speaker2023},
                Phi-4 Multimodal \cite{microsoftPhi4MiniTechnicalReport2025},
                Gemma 3n \cite{sansevieroIntroducingGemma3n2025},
                GPT4o-mini \cite{openaiGPT4oSystemCard2024},
                DeSTA2.5 \cite{luDeSTA25AudioGeneralPurposeLarge2025},
                Voxtral \cite{liuVoxtral2025},
                Whisper v3 Turbo \cite{radfordRobustSpeechRecognition2022},
                Kimi Audio \cite{kimiteamKimiAudioTechnicalReport2025},
                AudSemThinker (\cite{wijngaardAudSemThinkerEnhancingAudioLanguage2025}),
                Qwen2Audio \cite{chuQwen2AudioTechnicalReport2024},
                Qwen3 Omni \cite{yangQwen3TechnicalReport2025},
                MiMo Audio \cite{teamMiMoAudioAudioLanguage2025},
                Gemini 2.5 Flash \cite{comaniciGemini25Pushing2025},
                AudioFlamingo 3 \cite{goelAudioFlamingo32025},
                Qwen2.5 Omni \cite{xuQwen25OmniTechnicalReport2025},
                Gemini 3 Pro \cite{geminiteamNewEraIntelligence2025},
            },
            y dir=reverse,
            bar width=9pt,
            enlarge y limits=0.04,
            axis x line*=bottom,
            axis y line*=left,
            grid=major,
            grid style={dashed, gray!30},
            tick label style={font=\small},
            label style={font=\small},
            legend cell align={left},
            legend style={font=\small, at={(1.0,1.05)}, anchor=north east},
            legend image post style={xscale=0.7, yscale=0.7},
            bar shift auto=false,
            bar shift=0pt,
        ]

        \addlegendimage{area legend, fill=RoyalBlue, draw=none}
        \addlegendentry{Audio Understanding}
        \addlegendimage{area legend, fill=ForestGreen, draw=none}
        \addlegendentry{Speech-to-Text}
        \addlegendimage{area legend, fill=Brown, draw=none}
        \addlegendentry{Speaker Diarization}
        \addlegendimage{area legend, fill=Gray, draw=none}
        \addlegendentry{Auxiliary}

        \addplot+[xbar, fill=Gray, draw=none,
                  error bars/.cd, x dir=both, x explicit, error bar style={black}]
        coordinates {(-0.018462,1) +- (0.012952,0)};
        \addplot+[xbar, fill=Gray, draw=none,
                  error bars/.cd, x dir=both, x explicit, error bar style={black}]
        coordinates {(-0.000769,2) +- (0.011580,0)};
        \addplot+[xbar, fill=ForestGreen, draw=none,
                  error bars/.cd, x dir=both, x explicit, error bar style={black}]
        coordinates {(0.003000,3) +- (0.010755,0)};
        \addplot+[xbar, fill=ForestGreen, draw=none,
                  error bars/.cd, x dir=both, x explicit, error bar style={black}]
        coordinates {(0.004872,4) +- (0.007428,0)};
        \addplot+[xbar, fill=Brown, draw=none,
                  error bars/.cd, x dir=both, x explicit, error bar style={black}]
        coordinates {(0.015000,5) +- (0.010209,0)};
        \addplot+[xbar, fill=RoyalBlue, draw=none,
                  error bars/.cd, x dir=both, x explicit, error bar style={black}]
        coordinates {(0.023392,6) +- (0.019532,0)};
        \addplot+[xbar, fill=RoyalBlue, draw=none,
                  error bars/.cd, x dir=both, x explicit, error bar style={black}]
        coordinates {(0.025769,7) +- (0.010816,0)};
        \addplot+[xbar, fill=RoyalBlue, draw=none,
                  error bars/.cd, x dir=both, x explicit, error bar style={black}]
        coordinates {(0.027794,8) +- (0.021733,0)};
        \addplot+[xbar, fill=RoyalBlue, draw=none,
                  error bars/.cd, x dir=both, x explicit, error bar style={black}]
        coordinates {(0.030385,9) +- (0.026373,0)};
        \addplot+[xbar, fill=ForestGreen, draw=none,
                  error bars/.cd, x dir=both, x explicit, error bar style={black}]
        coordinates {(0.030667,10) +- (0.015259,0)};
        \addplot+[xbar, fill=ForestGreen, draw=none,
                  error bars/.cd, x dir=both, x explicit, error bar style={black}]
        coordinates {(0.035333,11) +- (0.011542,0)};
        \addplot+[xbar, fill=RoyalBlue, draw=none,
                  error bars/.cd, x dir=both, x explicit, error bar style={black}]
        coordinates {(0.036129,12) +- (0.021168,0)};
        \addplot+[xbar, fill=RoyalBlue, draw=none,
                  error bars/.cd, x dir=both, x explicit, error bar style={black}]
        coordinates {(0.051600,13) +- (0.013695,0)};
        \addplot+[xbar, fill=RoyalBlue, draw=none,
                  error bars/.cd, x dir=both, x explicit, error bar style={black}]
        coordinates {(0.058966,14) +- (0.024881,0)};
        \addplot+[xbar, fill=RoyalBlue, draw=none,
                  error bars/.cd, x dir=both, x explicit, error bar style={black}]
        coordinates {(0.070000,15) +- (0.028000,0)};
        \addplot+[xbar, fill=RoyalBlue, draw=none,
                  error bars/.cd, x dir=both, x explicit, error bar style={black}]
        coordinates {(0.080000,16) +- (0.029000,0)};
        \addplot+[xbar, fill=RoyalBlue, draw=none,
                  error bars/.cd, x dir=both, x explicit, error bar style={black}]
        coordinates {(0.089615,17) +- (0.025106,0)};
        \addplot+[xbar, fill=RoyalBlue, draw=none,
                  error bars/.cd, x dir=both, x explicit, error bar style={black}]
        coordinates {(0.092903,18) +- (0.022510,0)};
        \addplot+[xbar, fill=RoyalBlue, draw=none,
                  error bars/.cd, x dir=both, x explicit, error bar style={black}]
        coordinates {(0.098016,19) +- (0.026723,0)};
        \addplot+[xbar, fill=RoyalBlue, draw=none,
                  error bars/.cd, x dir=both, x explicit, error bar style={black}]
        coordinates {(0.100000,20) +- (0.023000,0)};

        \end{axis}
    \end{tikzpicture}
    }
    \caption{Shapley values per feature with standard error bars computed over 490 sampled configurations. Bars indicate the contribution of each audio tool to the overall system performance. Error bars are black.}
    \label{fig:agent:shapley_values}
\end{figure}

The best performing tools were incorporated into the configurations of this work (see Section \ref{ssec:agent:framework}), with two exceptions: Qwen2Audio \cite{chuQwen2AudioTechnicalReport2024} processes only 30 seconds of audio and the audios in the benchmarks are often longer, and AudSemThinker \cite{wijngaardAudSemThinkerEnhancingAudioLanguage2025} shares its architecture with Qwen2.5 Omni.

The Shapley value analysis reveals five key findings. First, general-purpose audio models (Qwen2.5 Omni: 0.098, AudioFlamingo 3: 0.093, Gemini 2.5 Flash: 0.090) contribute 2.5--3$\times$ more than speech-to-text models (Whisper v3 Turbo: 0.035), which shows that broader audio understanding models outweigh niche single-task models. Second, task-specific tools such as speaker diarization (0.015) provide modest value. Third, standalone benchmark performance does not predict tool effectiveness: AudioFlamingo 3 outperforms Qwen2.5 Omni on benchmarks \cite{goelAudioFlamingo32025} but achieves lower Shapley values (0.093 vs 0.098). Similarly, newer speech-to-text models (Parakeet TDT v3: 0.003, Granite Speech: 0.005) contribute less than older Whisper v3 Turbo (0.035) despite better transcription accuracy \cite{cohenNowWereTalking2025,saonGraniteSpeechOpenSourceSpeechAware2025}. Fourth, open-source tools match or exceed closed-source alternatives (AudioFlamingo 3: 0.093, Qwen2.5 Omni: 0.098 vs GPT4o: 0.028). Fifth, Gemini 3 Pro (0.100) is the highest contributor, while MiMo Audio (0.080) and Qwen3 Omni (0.070) are strong new additions that rank among the top tools.

\subsection{Discussion and Future Work} \label{ssec:agent:discussion_future_work}

The multi-hop consensus mechanism mitigates tool errors by querying multiple models and issuing targeted follow-up queries when outputs conflict, producing higher agreement rates than any single model. This approach implements a form of consensus aggregation where majority agreement across independent models provides a more reliable signal than any single model. Individual models may exhibit biases or hallucinations, but combining evidence from multiple sources reduces the impact of such errors.

The agent ablation reveals that tool-calling proficiency varies across models: the best closed-source agent (Opus 4.5: 0.752) outperforms the best open-source agent (Qwen 3 235B: 0.710) by 4.2 percentage points, while the gap between the weakest and strongest agent spans 13.6 percentage points. Orchestration quality depends on training methodology rather than model scale alone.

Shapley analysis shows that standalone benchmark performance does not predict tool effectiveness: general-purpose audio models contribute 2.5--3$\times$ more than specialized speech-to-text tools, and AudioFlamingo 3 outperforms Qwen2.5 Omni on benchmarks yet achieves lower Shapley values (0.093 vs 0.098). Future work should identify properties beyond accuracy that make models effective tools, including output format, response consistency, and instruction following.

A practical limitation is latency. Individual audio-language models process audio in 1--5 seconds per example, depending on model size. In contrast, AudioToolAgent requires 30 seconds to one minute per example because the framework coordinates multiple models sequentially, with the agent waiting for each tool's response before proceeding. Four strategies mitigate this overhead. First, the orchestrator can invoke multiple tool calls simultaneously when the tools are independent, reducing the number of sequential round-trips. Second, concurrent batching across examples allows the orchestrator to process multiple audio questions in parallel: while waiting for tool responses on one example, the orchestrator can initiate tool calls for other examples. Third, deploying multiple replicas of tools across GPUs enables load balancing and reduces queuing delays. Fourth, using API endpoints from external providers eliminates the need to run local GPU infrastructure while enabling parallel API calls at low cost per query. Because no training is involved, the total compute cost remains modest compared to fine-tuning or training new models. Future research should focus on developing learned tool selection policies and lower-latency orchestration strategies to reduce sequential dependencies.

The experiments tested the framework with web search integration using both DuckDuckGo API and Tavily's proprietary search API (see Figure \ref{fig:agent:shapley_values}). Ablation studies did not reveal consistent improvements, likely because the benchmarks focus on audio content requiring historical information and common knowledge facts. Web search integration may benefit real-world tasks where external knowledge retrieval is required beyond the audio content.

A next step is combining native multimodal processing with tool-calling capabilities, merging the efficiency of direct audio understanding with the extensibility of frameworks like AudioToolAgent.

\section{Conclusion}
\label{sec:agent:conclusion}

This paper introduced AudioToolAgent, a framework for multimodal audio understanding and reasoning where a central agent coordinates audio-language models as tools. The combination of Gemini 3 Pro, Qwen2.5 Omni, Audio Flamingo 3, and Whisper orchestrated by Gemini 3 Pro outperforms prior models on the MMAU, MMAR and MMAU-Pro benchmarks.

This framework combines the strengths of audio-language models and large language models. Ablation studies identified Gemini 3 Pro, Qwen2.5 Omni, and AudioFlamingo 3 as the highest-contributing audio tools across 20 evaluated tools. Among 12 LLMs tested as orchestrating agents, Claude Opus 4.5 and Gemini 3 Pro demonstrated the strongest performance, while open-source models like Qwen 3 235B achieved competitive results. This hybrid approach combines the audio processing of ALMs with the reasoning strengths of LLMs, creating a modular system in which tools and agents can be swapped independently.

The modular design enables integration of new tools and agents as they become available, without retraining. Future work includes developing learned tool selection policies, improving orchestration speed, and expanding the tool ecosystem to include audio retrieval and generation tools.

\section{Acknowledgments}
\label{sec:agent:acknowledgments}
This work was supported by the Dutch Research Council (NWO 406.20.GO.030 to Prof. Elia Formisano), the Dutch national e-infrastructure with the support of the SURF Cooperative using grant no. EINF-14224. GW and JJ acknowledge funding by the Federal Ministry of Education and Research of Germany (BMBF) under grant no. 01IS24085C (OPENHAFM), under the grant 16HPC117K (MINERVA) and under the grant no. 01IS22094B (WestAI - AI Service Center West), as well as funding by EU Horizon under grant no. 101214398 (ELLIOT) and co-funding by EU from EuroHPC Joint Undertaking programm under grant no. 101182737 (MINERVA) and from Digital Europe Programme under grant no. 101195233 (openEuroLLM).

\bibliographystyle{IEEEtran}
\bibliography{refs}

\end{document}